\documentclass[10pt,twocolumn,letterpaper]{article}

\usepackage{iccv}
\usepackage{times}
\usepackage{epsfig}
\usepackage{graphicx}
\usepackage{amsmath}
\usepackage{amssymb}
\usepackage[dvipsnames]{xcolor}

\usepackage[pagebackref=true,breaklinks=true,letterpaper=true,colorlinks,bookmarks=false]{hyperref}

\iccvfinalcopy 

\def\iccvPaperID{30} 

\ificcvfinal\fi
\begin{document}

\title{To Trust, or Not to Trust? A Study of Human Bias in Automated Video Interview Assessments}

\author{Chee Wee (Ben) Leong$^{1\dagger}$, Katrina Roohr$^{2\dagger}$, Vikram Ramanarayanan$^{1*}$, Michelle P. Martin-Raugh$^{2\dagger}$,\\ Harrison Kell$^{2\dagger}$, Rutuja Ubale$^{1*}$, Yao Qian$^{1*}$, Zydrune Mladineo$^{1\dagger}$, Laura McCulla$^{2\dagger}$\\
NLP, Speech, Dialogic \& Multimodal Research$^{1}$, Academic To Career Research$^{2}$\\
Princeton$^\dagger$, San Francisco$^*$, Educational Testing Service, USA\\
\and
}

\maketitle
\vspace{-4mm}

\begin{abstract}
Supervised systems require human labels for training. But, are humans themselves always impartial during the annotation process? We examine this question in the context of automated assessment of human behavioral tasks. Specifically, we investigate whether human ratings themselves can be trusted at their face value when scoring video-based structured interviews, and whether such ratings can impact machine learning models that use them as training data. We present preliminary empirical evidence that indicates there might be biases in such annotations, most of which are visual in nature.
\end{abstract}
\vspace{-4mm}
\section{Introduction and Related Work}

Structured interviews standardize the questions and/or evaluation methods used during the interview process (i.e., each interview has the same questions in the same order \cite{levashina2014structured}). As demonstrated by extensive meta-analytic evidence \cite{huffcutt2004impact,schmidt2004counterintuitive}, structured interviews consistently outperform unstructured interviews (i.e., an interview where questions and evaluations are non-standardized). They are particularly effective hiring tools that account for significant incremental validity in predicting job performance over and above other popular hiring methods such as personality and intelligence tests \cite{cortina2000incremental}. Despite their pervasiveness and effectiveness, interviews are susceptible to \textit{human biases}, a source of measurement error. Biases occur when interviewers collect or evaluate non-job-related information about applicants \cite{levashina2014structured}, such as sex, age, race, or attractiveness. Research shows that human interviewers tend to be less likely to hire individuals that are older \cite{morgeson2008review}, of a different race than themselves \cite{lin1992field}, perceived as unattractive \cite{hosoda2003effects}, or of a sex not stereotypically associated with a particular job (e.g., male nurse; \cite{davison2000sex}), among many other biasing factors. 
In this work, we differentiate human bias from \textit{human subjectivity}, where interviewers make different hiring recommendations for the same interviewee due to their different weightings of job-related strengths and weakness of the applicant. 
In spite of mounting evidence of human bias in industrial and organizational (I-O) psychology research, supervised machine learning approaches to automated scoring of video-based performance tasks (e.g. interview, presentations, public speaking, etc.) have largely focused on mitigating \textit{human subjectivity} by creating comprehensive rubrics \cite{schreiber2012development} to guide precise scoring of performance constructs \cite{chen2014towards}, enforcing the calibration of human raters prior to scoring \cite{chen2014towards,nguyen2014hire}, encouraging inter-rater discussions and reviews \cite{naim2015automated}, enlisting multiple raters \cite{nguyen2015would,ramanarayanan2015evaluating}, including both behavioral experts and layman as raters \cite{naim2015automated}, and averaging ratings \cite{naim2015automated,ramanarayanan2015evaluating,chen2017automated}. The work in \cite{nguyen2016hirability} attempted to collect demographic information such as age, gender and ethnicity as part of the data collection survey but did not explore their impact on automated scoring of the interviews. While the authors in \cite{naim2015automated} claimed additional Mechanical Turk raters were used as raters to remove bias from the actual interviewers due to their interaction with the interviewees, no effort was expended to further remove nor qualify the type of biases being investigated. 
In this paper, we present a study that motivates future efforts toward modeling fairness in automated video interview assessments. We first provide an overview of the video interview dataset used in our experiments. Next, we explain the annotation scheme used for generating the bias metadata vector, which is used to (1) construct a standalone model, and, (2) augment a multimodal model for structured interview performance prediction. Finally, we discuss some thoughts for future directions.

\section{Dataset}

Our video interview dataset is a corpus of monologic, structured video interviews collected online through Amazon Mechanical Turk from the authors in Chen et al. \cite{chen2017automated}. 
To our knowledge, this is the largest collection of structured interview responses simulating an actual hiring scenario (i.e. hiring for an entry-level office position). It comprises 260 human interviewees with a total recording time of 3784 minutes. Due to the dataset being collected ``in the wild'', there are videos where faces in dim lighting cannot be detected, or unexpected clips in audios. These represent around 10\% of the dataset and unfortunately have to be discarded since we failed to extract reliable multimodal features from them. Eventually, we have 1887  (interviewee, video response) datapoints after pre-processing.
All responses were collected from participants across the United States, and differed in gender, race, age, experience, etc. Recording conditions also varied in terms of devices, lighting, backgrounds, etc. All videos from interviewees are collected indoors. Chen et al. \cite{chen2017automated} developed a 7-point Likert rating scale (1 = Strongly Disagree, 7 = Strongly Agree) to score interview performance using overall hiring recommendation guidelines proposed in \cite{huffcutt2001identification}. During annotation, 5 human raters were asked to score each response using the Likert scale, and their averaged ratings are used as the ground-truth for the video. The same raters were used for scoring all the video interviewees. Subsequently, the authors used the median score of all ground-truths as a threshold to separate all video responses into HIGH (above-average) vs. LOW (below-average), and frame their experiments as a binary classification task using multimodal features to predict the outcome of each interviewee's performance. 
On the quality of human annotations, the authors reported an Intraclass correlation (ICC) of 0.79 (using the two-way random average measure of consistency), and R$_{mean}$ of 0.74, where R is the correlation coefficient of individual raters' scores to the averaged scores. Note that, for each interviewee, 8 different prompts (questions) were attempted. The response length is limited to 2 min/prompt.

\section{Coding Bias Metadata}
Research in I-O psychology has identified several categories of human biases relevant for our work, which we have listed here with the possible labels in parenthesis: SETTING (full room visible, partial room visible, only wall), WALL (blank, almost blank, with many items), GENDER (male, female), APPEARANCE (very unattractive, unattractive, average, attractive, very attractive), RACE (White, African American, Asian, Hispanic, Other), AGE (18-25, 26-35, 36-45, 46-55, 56-65, 65+), WEIGHT (very thin, thin, average, overweight, obese), FACIAL\_STIGMA (no, yes), ACCENT (American Typical, International, Domestic (e.g. Southern, Bostonian, etc.)). Two human raters trained in education and assessment research annotated each video interviewee independently on each of the categories using the appropriate label after going through an initial calibration. The two raters differed in race, accent, age when self-rated. Note that this bias metadata annotation effort is independent from the hiring recommendation annotation in \cite{chen2017automated}. 
Additionally, we hypothesize that physical backgrounds of interviewees \textit{may} induce rater bias. 
Hence, we introduce two categories coding for the visible environment (SETTING) and the physical state of the wall (WALL) behind the interviewee. Cohen's Kappa ($\kappa$)  between the 2 raters are reported for each category: SETTING (.62), WALL (.76), GENDER (.99), APPEARANCE (.49), RACE (.72), AGE (.69), WEIGHT (.70), FACIAL\_STIGMA (.13), ACCENT (.32). 
Where disagreement between the two raters occurs, we adopted a consistent rule for rounding  off to the integer coding for the nearest label for all ordinal categories (e.g. for APPEARANCE, if two labels coded are average (3) and attractive (4), then the average is 3.5, but the rounded label is attractive (4)). For non-ordinal categories (i.e. GENDER, RACE and ACCENT), we always arbitrate to rater 1's label where there is a disagreement for the sake of consistency. However, such disagreement cases account for only $\sim$2\% of the total non-ordinal label pairings. 
After the bias metadata vector for an interviewee is coded, the same vector is used for all 8 interview prompts answered by the same interviewee for experiments. We apply all labels in the bias metadata as a single feature set to construct models for predicting interviewee performance in a binary classification task (i.e. below-average or above-average). Since each of the 8 prompts are different, and responses across the 8 prompts are scored independently, each tuple (interviewee, video response) can be treated as single datapoint for experimentation. Of the 254 unique interviewees in the dataset, 152 ($\sim$60\%) have attained different classification scores across the 8 prompts, suggesting a potentially significant variance in the ability to handle different prompts even within an individual. We use stratified sampling in a 10-fold cross-validation applied at the (interviewee,prompt)-level to maintain distribution of the classes while sampling, and employ learners with proven effectiveness on relatively small-to-medium sized datasets using the scikit-learn toolkit \cite{pedregosa2011scikit}.

\section{Modeling structured interview through the exclusive use of bias metadata}

\begin{figure}[ht]
\vspace{-3mm}
\centering
\includegraphics[width=0.3\textwidth]{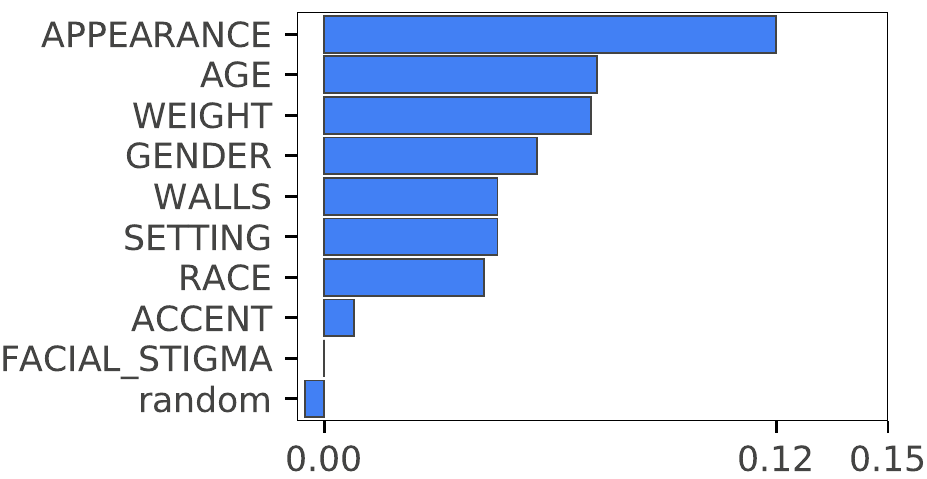}
\caption{Feature category importance weighting using RandomForestClassifier (n=500), 10-fold cross-validated}
\label{fig:rf_feature_importances}
\vspace{-3mm}
\end{figure}

\begin{figure}[ht]
\centering
\includegraphics[width=0.31\textwidth]{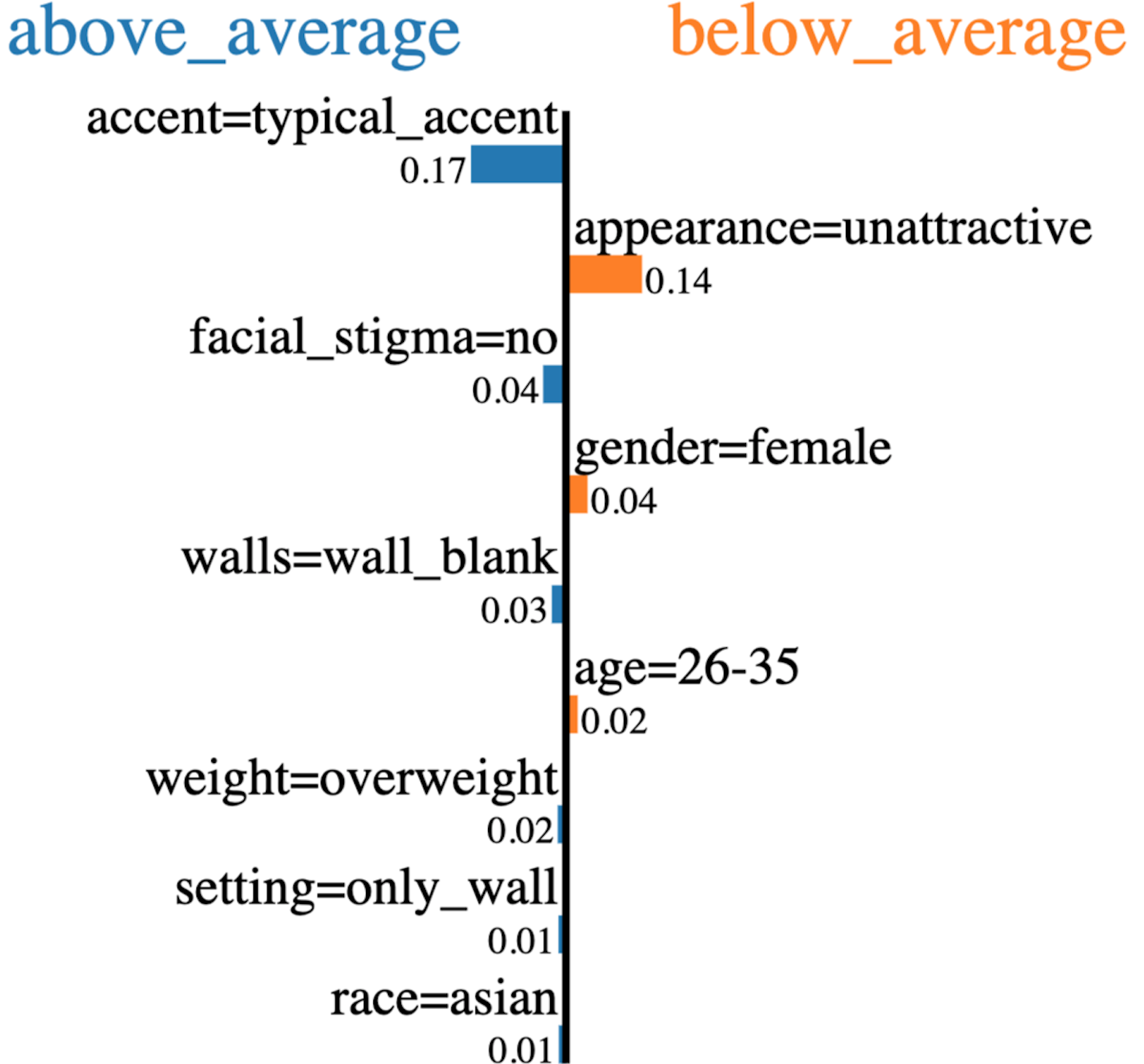}
\caption{Model interpretation using LIME applied to a prediction instance by RandomForestClassifier (n=500)}
\vspace{-8mm}
\label{fig:LIME_prediction}
\end{figure}

\noindent Prior to model building, each categorical label is converted into one-hot encodings fitted on the entire dataset of 1887 (interviewee, prompt) tuples, where only one label in each category per datapoint is activated, across all categories. Consequently, each label is transformed into a numerical value to facilitate experiments across a range of standardized learners. A stratified 10-fold cross-validation predictor that respects the class distribution of the train fold (i.e. a chance predictor), as well as a majority vote baseline, are used as the baselines. For the former, we used the StratifiedKFold function in \cite{pedregosa2011scikit}, where the proportion of classes in each train/test partition resembles that of the population, and all datapoints are \textit{randomized} and grouped into either the train or test partition with no repeats per fold. The results of the 10-fold cross-validation is shown in Table \ref{tab:augmented_eval}. Additionally, Figure \ref{fig:rf_feature_importances} shows the weight of each category (averaged over the 10-folds) used in the Random Forest model which scores the best performance (F$_{1}$=.765) in our combination experiments. The category weighting scheme is based on permutation importance \cite{rfpimp} instead of importance based on reduction in node impurity that is implemented in scikit-learn, which has reliability concerns \cite{strobl2007bias}. Note that category \textit{random} is a control with randomly generated values for ensuring validity in the feature importance computation: if a category has negative importance, removing it actually \textit{improves} performance.  Importance of the FACIAL\_STIGMA category (0) is currently inconclusive, due to overly skewed distributions (i.e. rater 1 and 2 annotated \textit{yes} on only 3.1\% and 1.9\% of the datapoints respectively). Otherwise, the most important label for modeling in each category are as follows: SETTING (\textit{partial\_room\_visible}), WALL (\textit{almost\_blank}), GENDER (\textit{female}), APPEARANCE (\textit{unattractive}), RACE (\textit{White}), AGE (\textit{26-35}), WEIGHT (\textit{overweight}),  ACCENT (\textit{Typical}). 

\begin{figure*}
\centering
\includegraphics[width=14cm]{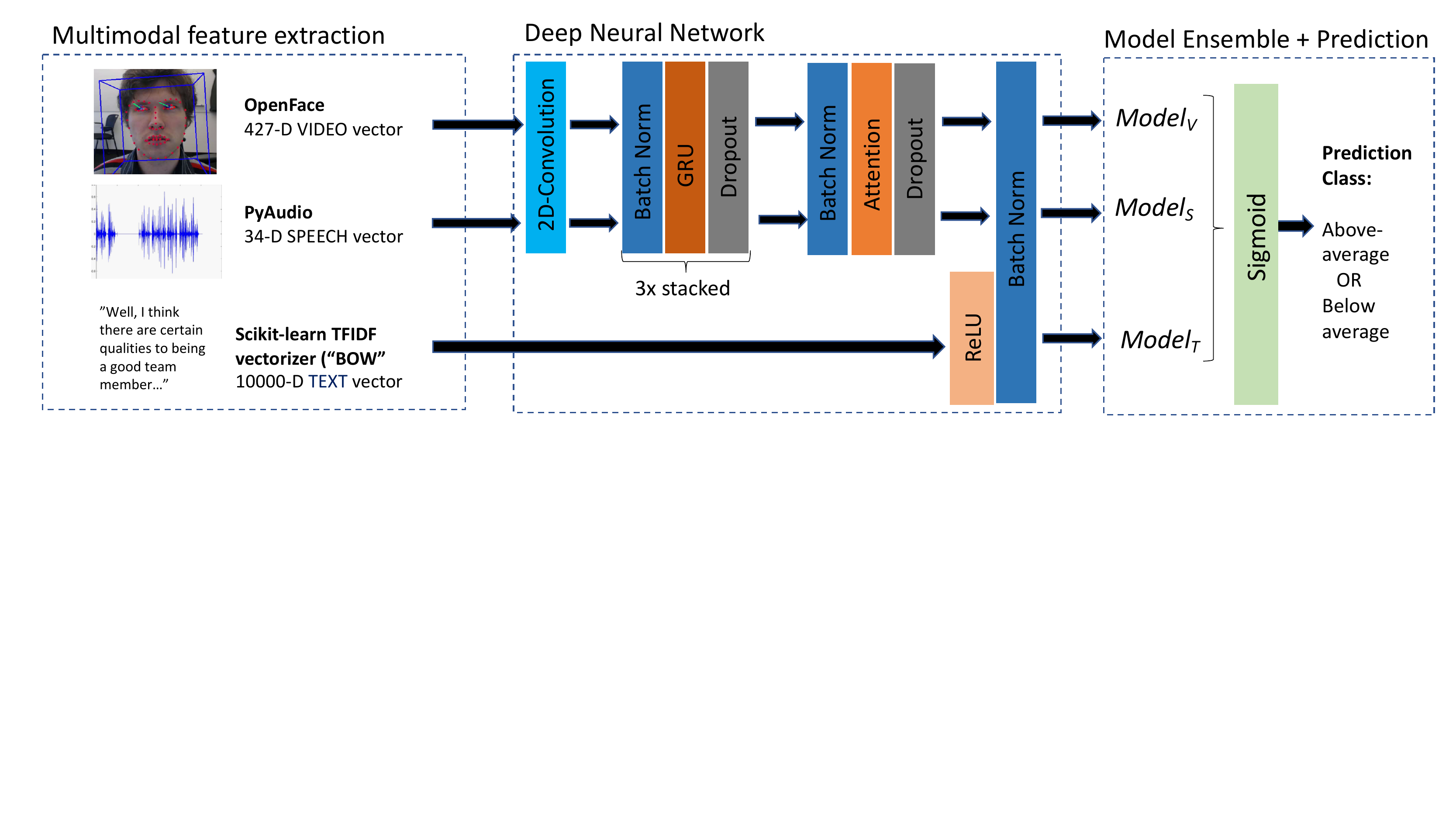}
\vspace{-40mm}
\caption{Multimodal DNN model for achieving the state-of-the-art in
\cite{chen2017automated}.}
\vspace{-3mm}
\label{fig:DNN_diagram}
\end{figure*}

Though encoded into numerical values for experimentation, the original labels (e.g. male, obese, unattractive, etc.) associated with each category are \textit{human-interpretable}, hence lending themselves to an explanation on whether any automated scoring model is behaving reasonably as measured against established I-O psychology findings. We take advantage of this phenomenon to further validate our hypothesis through application of the Local Interpretable Model-Agnostic (LIME)  \cite{ribeiro2016should} toolkit to our dataset, by examining datapoints where our model predictions are confident. For a targeted datapoint and its prediction, LIME perturbs inputs to other neighboring datapoints around it to learn an interpretable and high-fidelity, local model to help explain the prediction made by the input model. For instance, Figure \ref{fig:LIME_prediction} shows our Random Forest prediction $p(class="below-average")=1$ for a specific (interviewee, prompt) datapoint, with the contribution of each categorical label accounting for each candidate class prediction. Here, we note the interviewee being labeled with \textit{appearance=unattractive} adds a probability of 14\% to an unfavorable class prediction outcome without regard for multimodal features extracted. Simultaneously, the \textit{accent=typical\_accent} (typical American accent) adds a probability of 17\% to a favorable class prediction. This datapoint, and others we have examined, corroborate research findings in I-O psychology to a degree supported by the bias annotation agreement that biases \textit{might} influence interview outcomes. To the best of our knowledge, this work is the first that provides empirical linkage between I-O psychology and ML/AI modeling for structured video interviews. 


\section{Multimodal Model Augmentation}

We also experimented with multimodal model augmentation (i.e. an approach of augmenting a multimodal model with our bias metadata) to observe if we can further achieve model performance gains. For a fair comparison, we obtained the original feature set used in \cite{chen2017automated}, evaluate on the same training/testing partitions, but opted to use a deep neural network (DNN) for modeling the task due to its proven effectiveness for modeling other similar constructs (e.g. engagement \cite{dhall2018emotiw,yang2018deep} and emotions \cite{hazarika2018icon,vielzeuf2018occam}) of human behavior.

\begin{table}[tbh]
\small
\centering
\begin{tabular}{l|l|l|l}
\hline
 & \textbf{P} & \textbf{R} & \textbf{F$_{1}$}\\
\hline
\textbf{INDIVIDUAL LEARNERS} & & &\\
Logistic Regression & .581 & .567 & .572\\
Nearest Neighbor & .720 & .707 & .713\\
SVM (gamma=2, C=10)  & .742 & .756 & .748 \\
Decision Tree & .692 & .732 & .711\\
Random Forest (RF) (n=500)  & .737 & .750 & .742\\
Multilayer Perceptron (alpha=1) & .617 & .636 & .623\\
Multimodal DNN & .606 & \underline{.915} & .727\\
\hline
\textbf{COMBINATIONS WITH DNN} & & &\\
Multimodal DNN \& RF (AND) & \underline{.785$^*$} & .701$^*$ & .739$^*$\\
Multimodal DNN \& RF (Stacking) & .746$^*$ & .787$^*$ & \textbf{.765$^*$}\\
Multimodal DNN \& SVM (AND) & \textbf{.790$^*$} & .704$^*$ & .742$^*$\\
Multimodal DNN \& SVM (Stacking) & .755$^*$ & .773$^*$ & \underline{.763$^*$}\\
\hline
\textbf{BASELINES} & & &\\
Baseline (Stratified) & .512 & .552 & .531\\
Baseline (Majority vote) & .510 & \textbf{1.00} & .675\\
\hline
\end{tabular}
\caption{Mean 10-fold cross-validation of precision, recall and F$_{1}$ (true positive = above-average) of individual learners, selected combinations and baselines. All experiments are executed with the same random seed and random\_state for replicability. Bold and underline indicates first and second ranked results per column, while $^*$ indicates statistical significance at $p<.001$ over the Multimodal DNN system, using the Wilcoxon signed-rank test.}
\label{tab:augmented_eval}
\vspace{-4mm}
\end{table}

Given that we have only 1887 datapoints (1521/training, 366/testing), we train a deeper network but with dropouts to minimize overfitting. In our DNN model, a 2D convolution layer is applied to the time-series data i.e. video and audio to extract spatio-temporal features of interest within video/audio segments. Given the different sampling rates for video and audio feature extraction in \cite{chen2017automated}, and their different feature dimensions, we use different kernel sizes and filters (i.e. audio: 30 filters, each with kernel and strides (10,2); video: 20 filters, each with kernel and strides (15,16)) in order to generate a somewhat balanced representation of the two time-series modalities at the input level before sending them deeper into the network for abstraction. 
GRUs are used as our recurrent units which are favored over LSTMs for faster convergence in our experiment, with a final attention layer mechanism applied similar to the one in \cite{ubale2018exploring}. Our DNN model, shown in Figure \ref{fig:DNN_diagram}, is constructed and evaluated using Keras \cite{chollet2015keras} with a TensorFlow backend, and its hyperparameters are tuned using talos \cite{talos}. 
After 30 epochs in the same training partition, we achieved a best-performing model of \textbf{$F_{1}$=0.70} on the same test partition using the same multimodal feature set, which is competitive to the model performance of \textbf{$F_{1}$=0.66} achieved in \cite{chen2017automated} using SVM. With confirmation on the effectiveness of our DNN model, we retrained it with the same $F_{1}$ loss function and metric on each of the 10 train folds used in experiment 1. Next, we perform augmentation of the DNN model with the bias metadata using two approaches: (1) a simple, intuitive element-wise \textit{AND} condition between predictions made by the DNN model and the other learners, and (2) a stacked generalization method \cite{wolpert1992stacked} that uses the DNN prediction probabilities and combines it with the raw bias metadata vector before applying another learning algorithm to generate the final prediction. A justification for the latter approach is that the shorter but dense metadata bias vector may be masked by the much larger set of sparse, time-series modal features, hence combining them in the late fusion stage is more appropriate. Our results for the multimodal augmentation experiments are shown in Table \ref{tab:augmented_eval}. While using the logical AND augmentation generates a high-precision classifier, the stacked generalization approach is better at achieving overall model performance measured by $F_{1}$. 
\textbf{Inter-feature correlation}: As further investigation, we compute Cohen's Kappa ($\kappa$) between each of the bias metadata features and the multimodal DNN prediction output. The results indicate only very slight agreement between the biases and the prediction output, with most of the $\kappa$ centering at zero (GENDER has the highest $\kappa$ at .04). \textbf{Feature importance}: We again compute feature importances (Random Forest, n=500), this time using all available features. Consequently, the multimodal DNN prediction exclusively accounts for the largest importance weighting (.09), which is more than the weighting assumed by APPEARANCE (.06) that ranks second. These findings indicative that the multimodal DNN model by itself still accounts for a substantial variance in performance of a joint model. Bias metadata features have a non-negligible, combined importance weighting (.27), which is concerning if they are indeed construct-irrelevant. However, it could also be the case that some of these metadata features proxy for a construct-relevant latent trait that is not captured in the DNN features. We also note that these results reflect a closed-pool-of-subjects setting where the system will not see new subjects at test time, but only new interviews from existing subjects. Whether the findings generalize to a case with new interviewees, unseen at training time, is a question left for future work.



\vspace{-3mm}

\section{Conclusion}
\vspace{-1mm}
Despite having little correlation with construct-relevant multimodal features, the fact that demographic characteristics and other biasing variables have a non-negligible impact on modeling human behavioral tasks is worrisome, and poses implications for tasks such as the automated assessment of structured video interviews used in high-stakes employment settings where the decisions made using scoring models need to  be both fair and valid. Future work will further explore whether there is a causal relationship between these biases and human scores. If so, we will focus on debiasing techniques, possibly using avatars during human scoring that mirror an interviewee’s facial expressions without carrying over the visual biases, or modeling an individual human rater’s biases so that they can be statistically controlled for.

{\small
\bibliographystyle{ieee}
\bibliography{main}
}

\end{document}